\documentstyle[prl,epsf,aps]{revtex}

 \newcommand \be {\begin{equation}}
\newcommand \bea {\begin{eqnarray} \nonumber }
\newcommand \ee {\end{equation}}
\newcommand \eea {\end{eqnarray}}
 
 \newcommand \bi {\bibitem}
\newcommand \s {\sigma}
\newcommand \de {\delta}
\newcommand \De {\Delta}

\begin{document}
\twocolumn[\hsize\textwidth\columnwidth\hsize\csname@twocolumnfalse\endcsname
\draft      
%\preprint{MA/UC3M/09/96}
\title{Tempering simulations in the four dimensional $\pm J$ Ising spin
glass in a magnetic field}
\author{Marco Picco(*) and Felix Ritort(**)}
\address{(*) {\it LPTHE\/}\footnote{Laboratoire associ\'e No. 280 au CNRS}\\
       \it  Universit\'e Pierre et Marie Curie, PARIS VI\\
       \it Universit\'e Denis Diderot, PARIS VII\\
        Boite 126, Tour 16, 1$^{\it er}$ \'etage \\
        4 place Jussieu\\
        F-75252 Paris CEDEX 05, FRANCE\\
E-Mail: picco@lpthe.jussieu.fr\\
(**) Institute of Theoretical Physics\\ 
University of Amsterdam\\ Valckenierstraat 65\\ 
1018 XE Amsterdam (The Netherlands).\\ 
E-Mail: ritort@phys.uva.nl}

\date{\today}
\maketitle

\begin{abstract}
We study the four dimensional (4D) $\pm J$ Ising spin glass in a
magnetic field by using the simulated tempering method recently
introduced by Marinari and Parisi. We compute numerically the first four
moments of the order parameter probability distribution $P(q)$. We find
a finite cusp in the spin-glass susceptibility and strong tendency to
paramagnetic ordering at low temperatures. Assuming a well defined
transition we are able to bound its critical temperature.
\end{abstract} 

\vfill
%\pacs{05.30.-d, 64.60.Cn, 64.70.Pf, 75.10. Nr}
%{\bf \hfill cond-mat/9502045}
\twocolumn
\vskip.5pc] 
\narrowtext

%\vfill
%\newpage

%\baselineskip 6mm
\narrowtext
%123456789%123456789%123456789%123456789%123456789%123456789%123456789%12345678
Spin glasses are systems which deserve considerable theoretical interest
due to the interplay between randomness and frustration \cite{BOOKS}.
The role of the frustration in the statics and dynamics is essential to
understand the nature of the low temperature phase.  Despite great
progress during the last decade in the understanding of the mean-field
theory of spin glasses, a large number of topics are still poorly
understood. In particular, it is completely unclear which features of the
mean-field theory survive in finite dimensions. This problem has
recently received considerable attention \cite{NS,GUERRA,MAPARIRU} and
has become the cornerstone to validate the correct description of the
spin glass state.

The reason why this topic still remains open relies on the absence of a
convincing final theory for the spin glass state. Efforts to construct a
field theory of the glass state, based on the Parisi solution to the
mean-field theory, have been done mainly by De Dominicis, Kondor and
Temesvari \cite{DOKOTE}. Despite a large number of new results, a
clear answer to the finite dimensional issue is still missing.

After the Parisi solution to the mean-field theory, a new
phenomenological approach to the spin-glass state based on the
Migdal-Kadanoff renormalization group approach was proposed by McMillan
\cite{MILLAN}, Bray and Moore \cite{BM} and later on analyzed in detail
by Koper and Hilhorst \cite{KOHI} and Fisher and Huse \cite{FIHU}. In this
approach, the zero-temperature fixed point completely determines the
properties of the low temperature phase. This approach gave a description of
the spin glass state, now called the droplet model, where the
thermodynamics is determined by two Gibbs states (related by spin
inversion symmetry) plus a spectrum of excitations and corresponds to
the inversion of compact domains of finite size (droplets). This picture
of the spin-glass state lacks the most peculiar feature of the
mean-field theory, i.e. the coexistence of a large number of phases or
states in the spin-glass phase.

Recently, exact results have been obtained by Newman and Stein \cite{NS}
and also by Guerra \cite{GUERRA} on which features of the spin glass
state, present in the Sherrington-Kirkpatrick (SK) model \cite{SK},
survive in finite dimensions.  Numerical simulations are one of the few
confident tools that we can use to investigate this problem and clarify
the controversy \cite{NUMERICS}.  With the aid of numerical simulations,
two main questions in spin glasses have been addressed. The first one
concerns the low temperature behavior of the model in zero magnetic
field. The second one concerns the existence of the spin-glass
transition in a magnetic field similar to the one found by de Almeida
and Thouless in the mean-field case \cite{AT} (the so called AT line). A
clearcut answer to these questions would be very useful as a guide for
constructing a theory of the spin glass state in finite dimensions.
While the first problem has received considerable attention, very few
results have been obtained for the second one.

The purpose of this work is the study of the existence of spin-glass
state in a magnetic field. This work is the natural continuation of
previous numerical simulations done in the SK model in a magnetic
field, where the existence of a replica symmetry broken phase, as
predicted by Parisi \cite{PA}, was verified through the study of the
overlap probability distribution $P(q)$ \cite{PIRI}. In that work we
also studied the four dimensional (4D) $\pm J$ Ising spin glass in a
magnetic field in the low $T$ phase but did not find evidence for a
$P(q)$ of the mean-field type, even though we were not sure that
equilibrium was achieved for the largest sizes \footnote{We chose to
study 4D instead of 3D because the evidence in favor of a phase
transition at zero field is less obvious in this last case
\cite{OGI,MAPARI,KAYO,HUTANE}. Moreover the $4D$ model is easier to
thermalize than in $3D$.}. In order to investigate the existence of a
transition line in a magnetic field we have performed extensive tempering
Monte Carlo simulations in the $\pm J$ Ising spin glass in four
dimensions. 

{\em The model and the numerical algorithm}. 
We have considered the model described by the Hamiltonian,

\be
H=-\sum_{(i,j)}J_{ij}\s_i\s_j-h\sum_i\s_i
\label{eq1}
\ee

where the spin variables $\s_i$ take the values $\pm 1$, the $J_{ij}$ are
random discrete $\pm 1$ quenched variables and $h$ denotes the magnetic field.
The spins are located in the sites of a 4D cubic lattice
of size $L$ and $N=L^4$ sites with periodic boundary conditions.

In order to reach the maximum efficiency in the Monte Carlo simulations
we have used the tempering method introduced by Marinari and
Parisi \cite{MP}. This is a Monte Carlo method in which the temperature is a
dynamical variable and the system can change the temperature while always
being in thermal equilibrium. The system performs a random walk in temperature
in such a way that low temperature equiprobable configurations separated
by high energy barriers can be efficiently sampled. For a description
and details about this algorithm, the reader is referred to \cite{EM}.

In what follows we briefly describe the numerical procedure we have
followed. Samples are cooled down, at constant magnetic field $h$,
starting from the high temperature phase (above the critical temperature
at zero field $T_c\simeq 2.0$ \cite{BCPRPR}) down to $T=1.0$ and the
internal energy $e_{\beta}=\langle {\cal H}\rangle$ is estimated as a
function of $\beta$ for a selected set of $N_{\beta}$ different values
of $\beta$ ($N_{\beta}=50$ for the largest sizes). The separation
$\De\beta$ between the different values of $\beta$ is taken such that
the tails of the probability distributions of the energy for different
neighboring temperatures do superimpose. For sake of simplicity the
different values of $\beta$ were taken equidistant with $\De\beta=0.03$
for the largest sizes. It is important to note that all multicanonical
methods are expected to work if the {\em thermodynamic chaos} (to
be discussed below) is small. The weakness of chaotic effects in
temperature for finite sizes was numerically checked for the SK model
\cite{FR} as well as for 4D $\pm J$ Ising spin glasses \cite{FR2}.

%The different values of $\beta$ are separated by a quantity $\De\beta$
%such that configurations with energies between $e_{\beta}$ and
%$e_{\beta+\De\beta}$ are typical configurations of the equilibrium
%states at both temperatures $\beta$ and ${\beta+\De\beta}$.  

Starting from a random initial condition and an initial temperature
$\beta_r$, all the spins are sequentially updated at each Monte Carlo step
(MCS) and single spin flips are accepted with a probability given by
the heat bath algorithm.  After each MCS a change in temperature is
proposed $\beta_r\to\beta_{r+1}$ or $\beta_r\to\beta_{r-1}$, each with a
probability $1/2$. The change in temperature $\Delta\beta$ is accepted
with probability
$exp(-\Delta\beta(E(\s)-(e_{\beta}+e_{\beta+\De\beta})/2.))$. The spins
are again updated and the change of temperature is again proposed.  In
this way one is able to compute the equilibrium values of different
observables for all values of $\beta$. In order to increase the
statistics we have simulated 8 different replicas in parallel in a
multispin coding program.

Before presenting the numerical results, we will comment about 
chaotic effects in spin glasses and then explain how we choose the value of
the magnetic field for our simulations.

{\em Thermodynamic chaos in spin glasses}. One of the main properties
of spin glasses is the existence of chaotic effects when some external
parameter like the temperature or the magnetic field is changed
\cite{BM,KONDOR}. This feature is present in the mean-field approach
as well as in the droplet model at zero magnetic field. In the
framework of mean-field theory of spin glasses, the physical meaning of
thermodynamic chaos is rather intuitive. It is related to the fact
that small energy perturbations can redistribute the (small) free
energy differences of the many equilibrium states modifying completely
their equilibrium statistical weights. In the framework of droplet
models, energy perturbations can strongly modify spin correlations due
to the fractal nature of the droplet domain walls. We note that
thermodynamic chaos is a real effect of the spin glass
phase. According to the droplet model, the effect of a uniform magnetic
field is to suppress the spin glass phase, hence chaotic effects in
temperature disappear if the system becomes magnetized. The simplest way
to measure chaoticity in spin glasses is by defining the chaos
correlation length associated to the $q-q$ correlation function at
large spatial distances $x$ \cite{FIHU,KONDOR},

\be C_{chaos}(x)=\overline{\langle q_i\,q_{i+x}\rangle}\sim
x^{-\mu}\exp(-\frac{x}{\xi_c})
\label{eq2}
\ee where $q_i=\s_i\tau_i$ and $\s_i,\tau_i$ denote the spins of the
unperturbed (${\cal H}_0(\s)$) and perturbed system (${\cal
H}_p(\tau)$) respectively and the expectation value $<..>$ is taken
over the equilibrium Boltzmann distribution associated to the full
Hamiltonian ${\cal H}(\s,\tau)={\cal H}_0(\s)+{\cal H}_p(\tau)$.  $\mu$
is a positive exponent. The chaos correlation length ${\xi_c}$ gives
an estimate of the typical size of spatial regions which are similar
in the unperturbed and perturbed system. When the intensity of the
perturbation goes to zero, the chaos correlation length ${\xi_c}$
diverges and the divergence is related to the particular type of
perturbation.

For magnetic field perturbations, we know that chaotic effects
are quite strong \footnote{In contrast with chaotic effects in the
presence of temperature perturbations which are
small \cite{KONDOR,KOVE,FR}}. This effect sets a limit for the value of the
magnetic field that we can use in simulations. This is the most
relevant parameter in the simulations because it determines how
close we are to the $h=0$ spin-glass phase. The value of $h$ cannot be too
large otherwise, if a spin glass transition exists, it will be pushed
down to very low temperatures. Also it cannot be too small otherwise
the results are strongly affected by the $h=0$ spin-glass phase for
the finite sizes we have studied. The crossover between the $h=0$
behavior and the finite $h$ behavior depends on the chaos correlation
length $\xi_c(h)$ defined in eq.(\ref{eq2}). The value of the magnetic
field $h$ has to be chosen in such a way that $\xi_c(h)< L$ for the
explored lattice sizes but not too large as explained previously.  We have 
found that a good compromise is the value $h=0.4$ which yields a
macroscopic magnetization at low temperatures of order $0.15$. 
Then, we can estimate from independent Monte Carlo simulations (see
\cite{FR}) that $\xi_c\simeq 5$ for the lowest temperature $T=1$. We
expect that simulations for sizes above $L=5$ can yield convincing
results on the existence or absence of phase transition at this value of
the field. 

{\em Numerical results.} Simulations were performed for the following
sizes $L=3,5,7,9$ with $1000,325,120,130$ samples and
$N_{\beta}=20,40,50,50$ respectively ranging from $T_{min}=1.0$ up to
$T_{max}=2.5$ (or $T_{max}=3.0$ for the smallest sizes $L=3,5$). For the
largest size $L=9$ the number of temperatures was not large enough to
achieve equilibrium at low temperatures, hence we will show the data only
for temperatures above $T\simeq 1.6$ for that size.  In figure 1 we present
results for the magnetization $M=\frac{1}{N}\sum_i\s_i$ at different 
temperatures and sizes. Instead of plotting directly the magnetization
we plot the ratio $r(T,L)=\frac{MT}{h}$. This quantity (due to a local
gauge symmetry of the disorder \cite{OGI}) should be equal to 1 above
$T_c(h=0)\simeq 2.0$ in the limit of very small $h$. For finite $h$,
because of the divergence of the spin-glass susceptibility at zero
field, $r$ is smaller than 1 (at $T=2.5$ it is of order .7) but converges 
to 1 quite fast at high temperatures where the spin-glass susceptibility
vanishes like $\beta^3$. The important result which emerges in figure 1
is that below $T_c(h=0)$, $r$ is linear with $T$. Consequently the
magnetization is nearly constant in the low temperature phase. This
feature is also present in the mean-field theory and has been observed
in the 3D case \cite{CAPAPASO} as well as in field cooled experiments in
spin glasses \cite{BOOKS}.

\begin{figure}
\centerline{\epsfxsize=8cm\epsffile{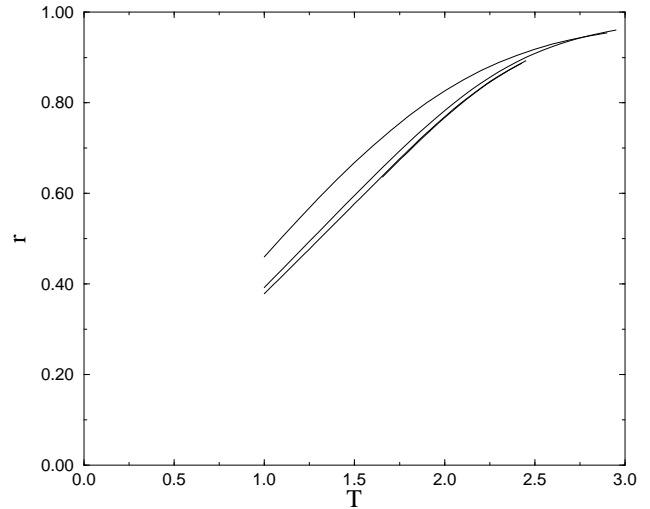}}
\caption{Parameter $r=\frac{MT}{h}$ as a function of temperature. From
top to bottom $L=3,5,7,9$. Data for $L=9$ is hardly distinguishable from
$L=7$.}
\end{figure}

More information about a possible phase transition can be obtained by
directly measuring the spin-glass order parameter $Q$ between two
different replicas $\lbrace\s,\tau\rbrace$ with the same set of
$J_{ij}$, $Q=\frac{1}{N}\sum_i\s_i\tau_i$ and its associated 
probability distribution,

\be
P_J(q)=\langle \de(q-Q)\rangle
\label{eq3}
\ee

where $\langle(\cdot)\rangle$ denotes the thermal Gibbs average. In
particular, for each sample we calculated the first four moments of
the distribution (\ref{eq3}).  We have computed the mean value, the
variance $X$, the skewness $Y$ and the kurtosis $Z$ of the
distribution $P(q)=\overline{P_J(q)}$ where $\overline{(\cdot)}$ means
average over the disorder. The skewness and the kurtosis are a measure
of the asymmetry and Gaussianity respectively of the overlap
distribution.  More precisely, if we define the following averages
$[f(q)]=\int dq f(q)P(q)$ we have,

\begin{eqnarray}
X=[(q-[q])^2]\\
Y=\frac{[(q-[q])^3]}{[(q-[q])^2]^{\frac{3}{2}}}\\
Z=\frac{1}{2}(3-\frac{[(q-[q])^4]}{[(q-[q])^2]^2})
\end{eqnarray}

In figure 2, we plot the mean value $[q]$ as a function of the
temperature for different sizes. Data for $L=9$ above $T\simeq 1.6$ is
nearly indistinguishable from data for $L=7$. As shown in figure 2, we
expect that $[q]$ converges to a value close to 1 at zero temperature
(but smaller than 1 if there is ground state degeneracy). The
cumulants $X,Y,Z$ give more information about a possible phase
transition. They are expected to vanish in the $L\to\infty$ limit in
the paramagnetic phase. Within an ordered phase of the mean-field
type, where several pure states contribute to the Gibbs average, we
expect $X,Y,Z$ to be finite.  In figure 3, 4, 5 we show $NX,Y,Z$
(where $N=L^4$) as a function of temperature for four different
lattice sizes $L=3,5,7,9$.

\begin{figure}
\centerline{\epsfxsize=8cm\epsffile{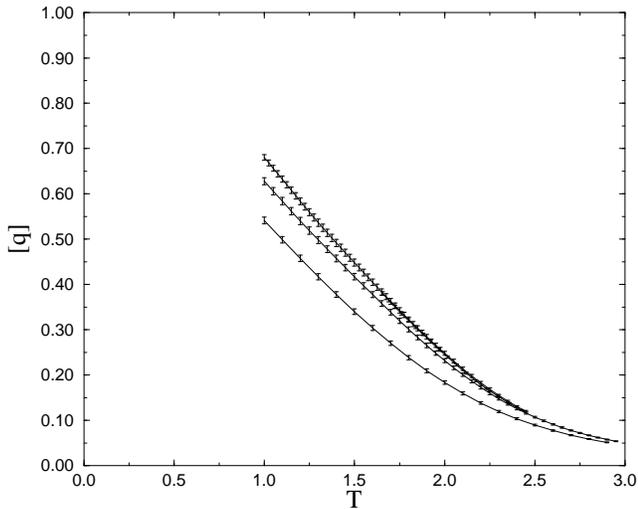}}
\caption{Mean value $[q]$ as a function of temperature. From
bottom to top $L=3,5,7,9$. Data for $L=9$ is hardly distinguishable from
$L=7$.}
\end{figure}

Figure 3 is quite interesting. We observe the presence of a cusp in the
spin-glass susceptibility for sizes $L=5,7$. This cusp moves to higher
temperatures as the size increases (for $L=3$ such a cusp is not
observed in the range of temperatures explored). The observation of
this effect already for $L=5,7$ reveals that it is a real trend of the
data and not a mere fluctuation. Unfortunately for $L=9$ we have not
been able to confirm or disprove this tendency (we have not succeeded
in thermalize $L=9$ at low temperatures).  According to the droplet
picture, the spin-glass susceptibility in the $\pm J$ model at zero
temperature should be positive at finite field due to the ground state
degeneracy (see the discussion in \cite{GRHE}). According to the
mean-field picture, the susceptibility should diverge below the AT
line. 

\begin{figure}
\centerline{\epsfxsize=8cm\epsffile{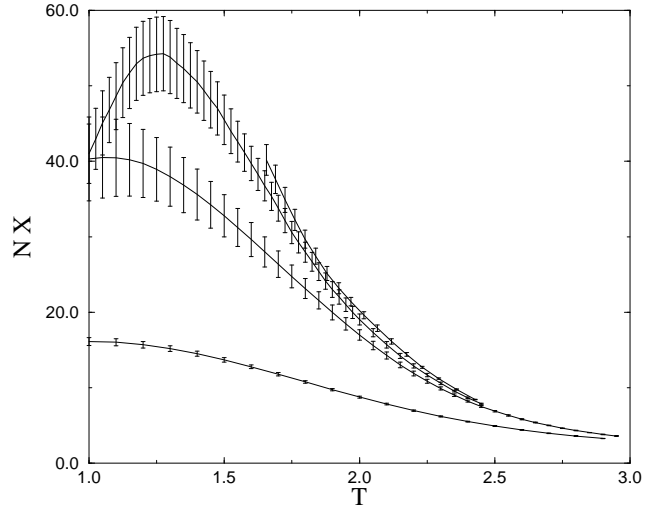}}
\caption{Spin glass susceptibility $NX$ as a function of temperature.
From bottom to top $L=3,5,7,9.$} 
\end{figure}

Figures 4 and 5 show the parameters $Y,Z$ for the different sizes.
What should be the manifestation of the existence of a second order
transition line? For large enough sizes, one expects the adimensional
quantities $Y,Z$ to scale like $Y\equiv {\hat Y}(L(T-T_c)^{\nu})$ (the
same for $Z$). Consequently they should display a crossing point for
different sizes at $T=T_c$ like happens at zero magnetic field
\cite{BCPRPR}. The lines in figures 4 and 5 corresponding to $L=5,7,9$
sizes have been indicated by full symbols in order to distinguish the
general trend of the data from the results for $L=3$. Figure 4 shows
that the skewness is negative for finite sizes and above $T\simeq 1.5$
it goes to zero when the size increases as expected in the
paramagnetic phase. The same tendency is also observed in figure 5 for
the kurtosis. It is interesting to note that the curves for $L=5,7$
for both the skewness and the kurtosis cross at the same temperature
$T\simeq 1.5$ which is an upper bound for of an hypothetical critical
temperature.  Unfortunately, we have not covered a large enough range
of sizes ($L=3$ is too small) in order to have clear evidence of such
a crossing point. Our results suggest the existence of paramagnetic
ordering at least above $T\simeq 1.5$ and we cannot exclude the
existence of crossing point and hence a phase transition below that
temperature.

\begin{figure}
\centerline{\epsfxsize=8cm\epsffile{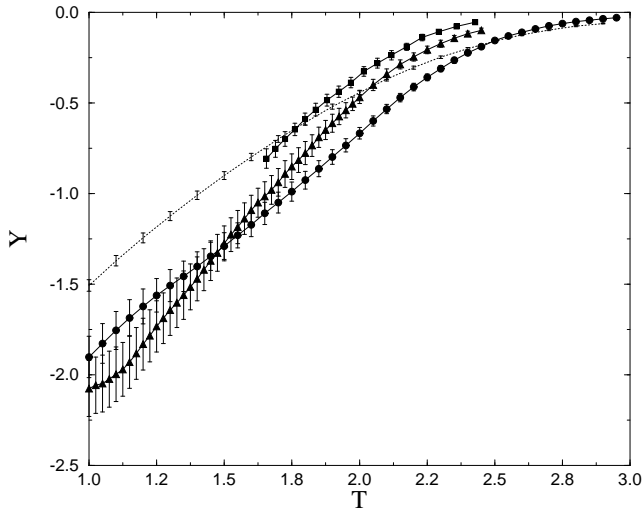}}
\caption{Skewness $Y$ as a function of temperature. Dotted line corresponds
to $L=3$, filled circles ($L=5$), filled triangles ($L=7$) and filled squares ($L=9$).} 
\end{figure}

\begin{figure}
\centerline{\epsfxsize=8cm\epsffile{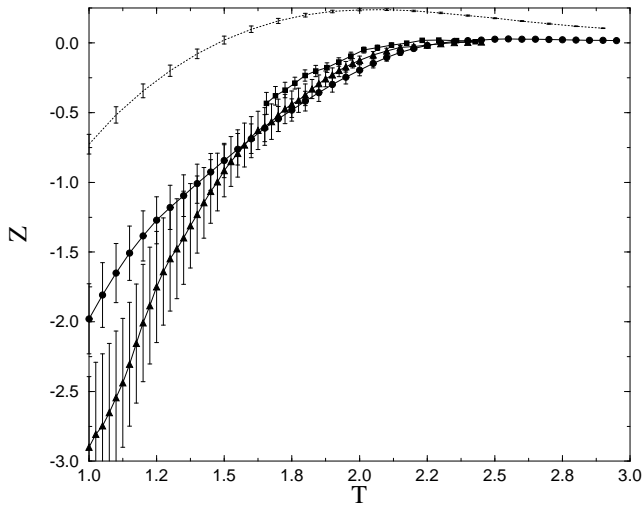}}
\caption{Kurtosis $Z$ as a function of temperature. Dotted line corresponds
to $L=3$, filled circles ($L=5$), filled triangles ($L=7$) and filled squares ($L=9$).} 
\end{figure}

From our data we reach the following three conclusions: 1) From figures
4 and 5 we can conclude that a phase transition, if existing, appears
below $T_c(h=0)\simeq 2.0$. Hence the transition temperature is
pushed down by the magnetic field.  2) We clearly observe a change of
behavior above $L=5$ where the trend of the skewness and kurtosis as a
function of the size changes. This crossover length is in agreement with
the estimated chaos correlation length $\xi_c\simeq 5$ for the value of
the magnetic field $h=0.4$ and should correspond to the size $L$ such
that the tail of the $P(q)$ extending down to negative values is
suppressed.  3) Most interestingly, figure 3 shows the existence of a
cusp which does not grow very fast with the size of the system and which
has to be appropriately interpreted, since it is in disagreement with the
mean-field picture. If the spin-glass susceptibility is finite at zero
temperature this cusp is then to be expected.  This result suggests that
if a transition exists, it should have non trivial finite-size effects.

A word of caution is essential at this point. Spin glasses are extremely
difficult to thermalize and this is probably the reason why small
progress has been done in the understanding of their equilibrium
properties.  The cusp observed in figure 3 can indeed be suppressed in
the presence of non equilibrium effects. The presence of this cusp was
not observed in \cite{BCPRPR} but in that case the magnetic field was
larger ($h=0.6$) and thermalization was achieved by simulated annealing
which is a less effective procedure than tempering.

It is important to note that the results we are presenting here are
probably seen in a very narrow range of fields. For fields larger than
$h=0.4$, the cusps in $X$ will move to lower temperatures and hence it is
difficult to see them numerically since tempering does not work efficiently
for very low temperatures. On the other hand, for smaller fields, the 
chaos correlation length would be larger and this would require to
simulate much larger lattices in order to start to see the trend of the
data.

Our data can then be interpreted in the framework of two scenarios: 1)
The transition is absent 2) The transition exists and the cusp in the
spin glass susceptibility is a finite size effect not incompatible
with a divergence in the thermodynamic limit. In this case the
transition should be above $T\simeq 1.25$ (where the cusp of $X$ in
fig. 3 for the largest size is located) and below $T\simeq 1.5$ (where
the crossing point for the skewness and the kurtosis is observed).  It
is difficult to go beyond such a conclusion and extremely careful work
(full thermalization in each sample is compulsory) has to be done in
order to discern among these possibilities. A similar work on the
Gaussian case (to avoid the ground state degeneracy) using the replica
exchange method recently proposed by Hukushima, Takayama and Nemoto
\cite{HUTANE} would be welcome in order to check the main conclusions
of this work.

{\bf Acknowledgments.} We acknowledge stimulating discussions with
S. Franz and also G. Parisi for a careful reading of the
manuscript. One of us (F.R.) is grateful to E.N.S in Paris for its
kind hospitality where part of this work was done. The work by F.R has
been supported by FOM under contract FOM-67596 (The Netherlands).

\vfill

\end{document}